\documentclass[iop]{emulateapj}
\usepackage{color}
\usepackage{graphicx}
\usepackage{amsmath}

\DeclareGraphicsExtensions{.pdf,.ps}

\begin{document}
\bibliographystyle{astroads}

\title{A New Model for the Radio Emission from SN 1994I and an Associated Search for Radio Transients in M51}

\author{Kate D. Alexander, Alicia M. Soderberg, Laura B. Chomiuk\altaffilmark{1}}
\affil{Harvard-Smithsonian Center for Astrophysics, 60 Garden Street, Cambridge, MA 02138}
\email{kalexander@cfa.harvard.edu}
\altaffiltext{1}{Department of Physics and Astronomy, Michigan State University, 567 Wilson Road, East Lansing, MI 48824} 

\submitted{Accepted to the Astrophysical Journal}

\begin{abstract}
\medskip

We revisit the exquisite archival radio data for the Type Ic supernova SN 1994I and present a revised model for the SN radio emission and a pilot study that aims to constrain the rate of C-band radio transients within the face-on host galaxy, M51 (NGC 5194).  We find that the temporal and spectral evolution of the SN\,1994I radio emission are well fit by a synchrotron self-absorption model and use this to estimate physical parameters. We compute a pre-explosion mass loss rate of $\dot{M}=3.0 \times 10^{-5} M_{\odot}$ yr$^{-1}$ for the progenitor, consistent with those observed from galactic Wolf-Rayet stars. Our model makes different assumptions for the dynamical model for the shockwave interaction than the model previously published by \cite{weil11}, but our $\dot{M}$ is consistent with theirs to within errors and assumptions. Drawing from a subset of the archival radio observations from the Very Large Array collected for the monitoring of SN\,1994I, we conduct a  pilot study to search for previously-unidentified transients. Data were primarily taken at a frequency of 4.9 GHz and are logarithmic in cadence, enabling sensitivity to transients with variability timescales ranging from days to months. We find no new transient detections in 31 epochs of data, allowing us to place a $2\sigma$ upper limit of 17 deg$^{-2}$ for the source density of radio transients above 0.5 mJy ($L\ga4\times10^{25}$ erg s$^{-1}$ Hz$^{-1}$ at the distance of M51). This study highlights the feasibility of utilizing archival high-cadence radio studies of supernova host galaxies to place constraints on the radio transient rate as a function of luminosity in the local Universe.

\smallskip
\end{abstract}


\section{Introduction}

Twenty years after its discovery, SN 1994I retains its importance as
one of the best-studied Type Ic supernovae (SNe Ic) to date (e.g.,
\citealt{filip95,rich96,mil99,weil11,imm02}). SNe Ic are
distinguished from other SNe by the lack of strong hydrogen or
helium lines in their optical spectra \citep{filip97}. They are
often grouped with Type Ib SNe (showing detectable helium features) as
a broader class of hydrogen-poor core-collapse explosions and we refer
to them collectively as SNe Ib/c. The progenitors of SNe Ib/c have not 
yet been directly observed (\citealt{smar09}, \citealt{eld13}) but are
commonly thought to be massive stars that have lost their outer
envelopes prior to explosion due to stellar winds or the
interaction with a companion.  Two primary classes of progenitor candidates
have been suggested for SNe Ib/c: (i) Wolf-Rayet stars \citep{woos95} and (ii) low mass
helium stars with close binary companions undergoing mass transfer
\citep{wel99,yoon10}.


SN\,1994I is located within the nearby spiral galaxy M51 (NGC 5194, the Whirlpool Galaxy) at a distance of 8.4 $\pm$ 0.6 Mpc \citep{feld97}. 
The event was discovered on 1994 April 2 UT by four independent sets of observers \citep{puc94}. Pre-discovery optical imaging of the field obtained prior to April 2 constrain the explosion date to $\sim$ a few days prior to the first detection on 1994 March 31.43 UT \citep{rich96}.  The relative proximity of SN\,1994I enabled astronomers to obtain detailed and densely sampled observations across the electromagnetic spectrum (e.g. \citealt{filip95,rich96,mil99,imm02,weil11}). These data enabled constraints on the progenitor models for Type Ib/c SNe and still form a standard template to which observations of other Type Ic SNe are compared \citep{fish97,branch07}.  This paper focuses on a re-analysis of the archival radio data of M51 originally obtained to monitor the non-thermal emission from SN\, 1994I \citep{weil11}.  

Radio observations of SNe provide a unique probe of the non-thermal synchrotron emission generated as the SN blastwave passes through the circumstellar environment, shocking and accelerating particles ejected prior to the explosion by the wind from the progenitor star \citep{chev96}. At a given radio frequency the radio SN luminosity rises to a peak and subsequently declines over a period of several months, thus dictating the logarithmic cadence of the follow-up observations. The physical properties of the shockwave and the explosion environment, including the temporal evolution of the shockwave radius $R$ and its time-averaged velocity $\bar{v}$, the magnetic field strength $B$, the total internal energy of the radio emitting circumstellar material $E$, and the pre-explosion mass loss rate $\dot{M}$, may be derived through temporal and spectral modeling of the non-thermal emission \citep{chev98} (see Section \ref{sec:94i}).  A detailed analysis of this radio dataset by \citealt{weil11} resulted in a mass loss rate estimate for the progenitor of SN\,1994I of $\dot{M}\approx 1.8 \times 10^{-7} (v_s/$20,000 km s$^{-1})^{1.5}(v_w/10$ km s$^{-1})(T_e/$20,000 km s$^{-1}) M_{\sun}$ yr$^{-1}$, where $v_s$ is the supernova shock speed, $v_w$ is the assumed wind speed of the progenitor, and $T_e$ is the electron temperature.  Here we report the results from our independent analysis and modeling of the radio SN emission and report estimates for each of the physical parameters ($R, \bar{v}, B, E,$ and $\dot{M}$) utilizing our own distinct methods. We also compute the brightness temperature $T_B$ and discuss its evolution.

Additionally, we utilize the densely-sampled radio observations of the M51 field obtained through the SN\,1994I follow-up campaign to search for other transients situated within the host galaxy (Section \ref{sec:trans}). While not as extensive as other radio transient studies, this pilot study demonstrates that radio transient science may be extracted from archival SN monitoring programs with modest additional effort.  A variety of radio transients have been observed and/or predicted, including radio supernovae \citep{weil86, berg03}, radio pulsars \citep{hew68, john03}, afterglows associated with gamma ray bursts \citep{frail97, levin02}, radio flares from brown dwarfs or extrasolar planets \citep{berg01, zar07}, X-ray binaries \citep{han01}, and tidal disruption events \citep{zau13}. Radio transients have the potential to exist at all radio wavelengths and can occur on timescales ranging from nanoseconds to years \citep{laz09}.  The exploration of this vast parameter space has only just begun (see \citealt{fend11} or \citealt{frail12} for a review). 

A significant fraction of radio transients observed to date were either first discovered at other wavelengths or discovered serendipitously in the course of making other radio observations. A number of blind imaging surveys have also been undertaken to search for radio transients, e.g., \cite{levin02} and \cite{gal06} found potential transient objects by comparing the FIRST and NVSS 1.4 GHz catalogs, covering an area of 2400 deg$^2$ to a sensitivity of 6 mJy, and \cite{ban11} detected 15 transient objects in a 2776 deg$^2$ survey of the Southern sky at 843 MHz.  More recently, \cite{bel14} found no transients in a 1430 deg$^2$ survey at 154 MHz, allowing them to set an upper limit of $7.5 \times 10^{-5}$ deg$^{-2}$  for the surface density of sources with flux densities $>$5.5 Jy. However, most radio imaging surveys are limited to single fields and due to data availability most single-field searches for radio transient objects have been limited to calibration fields, as these are the most frequently observed. For example, \cite{bow07} performed an extensive study of VLA data, using 944 epochs of a single field at two frequencies (5 and 8.4 GHz) taken over a 22-year period and finding 10 candidate transient objects with characteristic timescales between 7 days and 2 months. Additional work by \cite{frail12} eliminated some of these candidates and reduced the significance of others, emphasizing the importance of careful data reduction techniques and the availability of multi-wavelength data for follow-up observations. The nature of the remaining transients remains unknown. 

While calibration fields are useful for transient surveys due to the fact that they are observed repeatedly \citep{bow07}, they are non-ideal because they are typically tuned to a single bright calibration source. While they are just as likely as any other field to contain very luminous transients at large distances, a field containing a nearby galaxy will allow us to probe the lower luminosity population. This consideration makes the VLA dataset for M51 an ideal candidate for a targeted radio transient search since the distance to M51 ($d\approx 8.4 \pm 0.6$ Mpc) and its line-of-sight orientation (face-on) are well constrained, so we may make the reasonable assumption that super-posed transients originate within the galaxy. Since the distance is known, we can express the resulting constraints on transient rates in terms of luminosity rather than flux. This allows us to constrain the occurrence rates of specific types of transients of known luminosities within M51. We note that supernova ``factories'' such as M51 are popular targets for radio observations as several supernovae (SN 1994I, SN 2005cs, and SN 2011dh) have occurred within the galaxy in recent decades.  As a result, the archival dataset contains over 100 hours of public VLA observations of M51 at multiple wavelengths and frequencies spanning a period of nearly 30 years.  We focused our transient search around observations of SN 1994I taken in the months following its discovery since they are closely spaced and of similar depth, thereby facilitating a comparison between epochs.  This pilot study verifies the feasibility of using archival VLA observations of nearby galaxies for transient searches and allows us to place an upper limit on the rate of radio transients in local galaxies. Our technique could be expanded to archival radio observations of other nearby galaxies, such as M82, where a similar SN monitoring program has already led to the serendipitous discovery of a transient thought to be an unusual microquasar \citep{jos11}. SN 2008iz in M82 was also discovered serendipitously using radio observations originally collected for other purposes \citep{brun09,brun09b}.

In the next section, we describe the dataset used for our analysis and our data reduction techniques. We report the results of the SN study in \S\ref{sec:94i}, and the results of the targeted radio transient search in M51 in \S\ref{sec:trans}.  In \S\ref{sec:94i} and \S\ref{sec:conc} we compare our results with those in the literature. 

\section{Observations and Analysis}\label{sec:obs}

\subsection{Dataset and Analysis}

\begin{figure*}
\epsscale{0.9}
\plotone{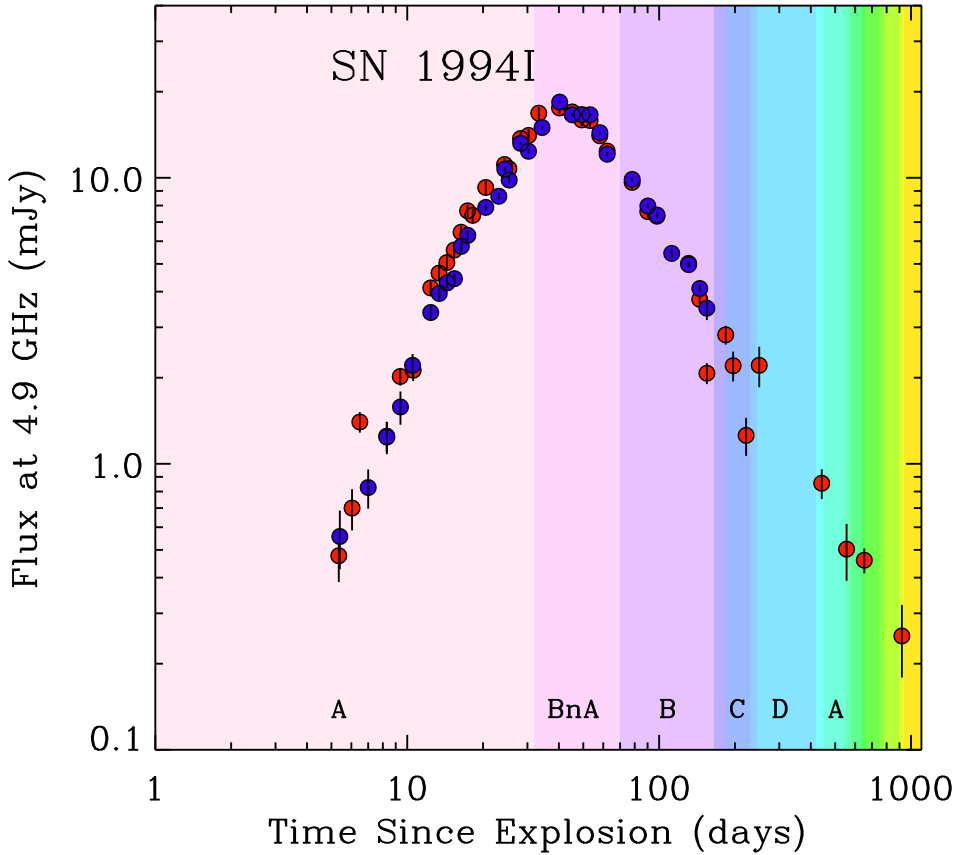}
\caption{\small C-band (4.9 GHz) observations of SN 1994I taken over a period of several years following its discovery. Our points are shown in blue and are compared with those of \cite{weil11} in red. For this figure we assume an explosion date of March 31.0 UT. The configuration of the VLA at the time of each observation is shown with shaded bands.}
\label{fluxes}
\end{figure*}

We use all available 4.9 GHz archival radio data for the field of SN\,1994I obtained with the Very Large Array (VLA)  spanning four configurations over several years following the SN explosion. 

The data set includes 3C286 as the flux density calibrator and a primary phase calibrator of J1418+546 with an approximate flux density of 0.97 Jy during the time period of the observations.   Intensive VLA monitoring of SN 1994I began three days after its optical discovery, providing one of the first opportunities to study the early radio emission from a SN Ib/c \citep{weil11}. As observations taken at frequencies $\sim 1$ GHz in the compact configurations were found to be dominated by diffuse emission from M51, we restricted our transient search to C-band (4.9 GHz) observations taken between 1994 April 5 and 1994 November 7 to maximize the combination of sensitivity, resolution, and minimal host galaxy contamination.  These observations span configurations A, BnA, and B.

We independently reduced 31 epochs of data taken in the A, BnA, and B configurations to identify transient candidates systematically in the dataset at 4.9 GHz and used them to generate maps of the M51 field. Most of the data were obtained under program AS0525 (PI Sramek) as part of a long-term monitoring project of SN\,1994I \citep{weil11}. In addition to the 4.9 GHz observations, the dataset also includes observations spanning 1.4-22.5 GHz, allowing us to investigate possible transient candidates at other frequencies and obtain a full radio spectrum.  All observations were conducted in dual polarization mode and have total integration times of 9-20 minutes. Each observation has a total bandwidth of 100 MHz, composed of two 50 MHz-wide IFs bracketing the central frequency for each band given above. 

The data were reduced using standard routines with the software package AIPS (the Astronomical Image Processing System) for calibration purposes, and coupled with our pipeline routine in order to produce final maps of the field. 
We sought to produce radio images covering the entire primary beam (extending out to at least the half power beam width). No primary beam correction was applied, as the full M51 galaxy fits within the primary beam and SN 1994I is located at the center of the images. The area imaged ranged from 11 arcmin by 11 arcmin for the A configuration epochs to 13.6 arcmin by 13.6 arcmin for the B configuration epochs.  Further details of the images used in the transient search are given in Table \ref{tab_obs}. 
We fit a Gaussian to the SN\,1994I radio counterpart in each map and extract the total integrated flux density. As a cross-check, we compare our measurements with those of \cite{weil11} in Figure \ref{fluxes}. We find that the two sets of fluxes agree well, demonstrating the validity of our pipelined analysis and mapping technique at 4.9 GHz.

\begin{table*}
\center{
\caption{VLA Images Used in Transient Search}
\label{tab_obs}
{\small
\begin{tabular}{ccccc}
\hline
\hline\noalign{\smallskip}
Configuration    & Number of &  Image Dimension & Clean Beam  & Typical rms Noise\\
   & Epochs & (arcmin) & (arcsec) & ($\mu$Jy beam$^{-1}$) \\
\hline\noalign{\smallskip}
A    & 18 & 11.0 & 0.35 x 0.27 & 90 \\
BnA  & 6  & 11.4 & 1.2 x 0.48 & 60 \\
B    & 7  & 13.6  & 1.4 x 1.1 & 100 \\
\hline\noalign{\smallskip}
\end{tabular}
}
}
\end{table*}

\subsection{Transient Search}

We used the AIPS task SAD (Search and Destroy) to identify all radio objects within each image consistent with a Gaussian fit.   We ran two iterations, with flux density thresholds of 10$\sigma$ and 5-6$\sigma$.  The transient detection threshold was set depending on the quality of the data in each image. The resulting mean central detection threshold of the images was 430 $\mu$Jy. Assuming Gaussian noise, we predict no more than one false positive in our sample of 31 epochs for this cutoff.  However, some calibrations were imperfect, due to non-ideal weather conditions and limited UV coverage, resulting in a number of detections generated by the resulting imaging defects.  These were removed from our source catalogs for each image after a visual inspection of the data. The AIPS task JMFIT was then used to fit the identified objects in each epoch, including SN 1994I and the galactic nucleus of M51.\\

\section{Modeling of SN 1994I}
\label{sec:94i}
 
 \subsection{Synchrotron Self-Absorption Model}
 
\begin{table*}
\center{
\caption{Physical Properties of SN 1994I. Based on fits of our synchrotron self-absorption-only model (Section \ref{sec:94i}) to \cite{weil11}'s reported fluxes assuming an explosion date of March 30.0 UT.}
\label{tab3}
{\small
\begin{tabular}{cccccccccc}
\hline
\hline\noalign{\smallskip}
Age   & $\nu_{p}$ & $F_{\nu_p}$ & R  & $\bar{v}$  & B  & E$_\epsilon$  & $\dot{M}$ & $T_B$ at 4.9 GHz &  $\chi^2$ per degree\\
(days)  & (GHz) & (mJy)       &   $(10^{15}$ cm)$^a$   &  ($c = 1$)$^b$  & (G)$^c$ & ($10^{45} $erg)$^d$  & ($10^{-5}M_{\sun}$ yr$^{-1}$)$^e$   & ($10^{10}$ K)$^g$ & of freedom\\
\hline\noalign{\smallskip}
10.125	& 22.5	& 14.80 & 2.39 & 0.0912	& 2.38	& 6.47 & 4.8 & 2.63 & 1.68\\
11.5$^f$ 	& 13.94 	& 19.44 & 4.39 & 0.147 	& 1.43 	& 14.5 & 2.3 & 3.38 & 0.648\\
12.675	& 15.0	& 18.65 & 4.00 & 0.121	& 1.549	& 12.8 & 3.2 & 3.25 & 2.14\\
21.575	& 8.4		& 18.35 & 7.09 & 0.127	& 0.869	& 22.5 & 2.9 & 4.34 & 8.39\\
34.3 		& 4.9 	& 17.25 & 11.8 & 0.133 	& 0.510 	& 35.8 & 2.5 & 3.58 & 7.46\\
35.2$^f$  	& 4.61 	& 14.85 & 11.7 & 0.128 	& 0.488 	& 31.7 & 2.4 & 3.21 & 3.63\\
101.5 	& 1.5 	& 9.875 & 29.6 & 0.113 	& 0.166 	& 59.5 & 2.0 & 0.157 & 3.07\\
131.9$^f$ & 1.525 	& 9.915 & 29.2 & 0.0854 	& 0.168 	& 58.8 & 4.1 & 0.165 & 0.470\\
\hline\noalign{\smallskip}
\end{tabular}
\parbox{6.5in}{\small
$^a$ Radius of blastwave in cm \\
$^b$ Time-averaged expansion velocity in units where $c = 1$\\
$^c$ Post-shock magnetic field strength in gauss\\
$^d$ Total internal energy of the blastwave in ergs \\
$^e$ Pre-explosion mass loss rate of the progenitor in solar masses per year, assuming $v_w \approx 1000$ km/s (typical for a Wolf-Rayet star)\\
$^f$ Parameters derived from the best-fit model for one of the spectra shown in Figure \ref{spec}. Otherwise, parameters are derived from the best-fit model for one of the light curves shown in Figure \ref{mods}. \\
$^g$ Brightness temperature of the emitting region at 4.9 GHz
}
}
}
\end{table*}

In order to extract the physical properties of the SN\,1994I explosion, its environment, and its progenitor star, we fit the observed radio flux density values of the SN with a model outlined in \cite{chev98} and \cite{chev06}. Our model assumes only synchrotron self-absorption, a simplification that allows the data to constrain the model sufficiently to self-consistently solve for the blastwave velocity $\bar{v}$ and the pre-explosion mass loss rate $\dot{M}$ at multiple epochs. In addition to directly computing the expansion velocity and mass loss rate, we also report the radius of the blastwave, the post-shock magnetic field strength, and the total internal energy of the radio-emitting ejecta at several times. This allows us to probe possible temporal variations in the physical parameters and make accurate inferences about the physical properties of the SN environment and its progenitor. 

In our model, the radio emission is explained as synchrotron emission from an interaction region between the supernova ejecta and material ejected from the progenitor star in a pre-explosion stellar wind. The rising part of the light curve is explained by synchrotron self-absorption. At the peak, the synchrotron emission transitions from optically thick to optically thin. It then fades as the blastwave decelerates and the density of the circumstellar material (CSM) decreases. The observed flux $F_\nu$ at radio frequency $\nu$ as a function of the time $t$ elapsed since the explosion is given by \citep{chev98}:
\begin{equation}
\begin{split}
\label{eq1}
\frac{F_\nu(t)}{F_{\nu_p}(t_p)} = 1.582\left(\frac{t}{t_p}\right)^a\left(\frac{\nu}{\nu_p}\right)^{5/2}\quad\quad \quad\quad  \quad\quad\quad\quad \\
\times \left\{1-\exp\left[-\left(\frac{t}{t_p}\right)^{-(a+b)}\left(\frac{\nu}{\nu_p}\right)^{-(p+4)/2}\right]\right\}.
\end{split}
\end{equation}
Here, $F_{\nu_p}$ is the peak flux at frequency $\nu_p$, $t_p$ is the time of peak flux at frequency $\nu_p$, $a=2m+0.5$, and $b=(p+5-6m)/2$, where $p$ is the energy spectral index of the relativistic electrons generating the synchrotron emission and $m$ is related to the outer density profile of the SN ejecta. The electrons are assumed to have a power-law distribution $N(E)=N_0E^{-p}$ and $m=(n-3)/(n-2)$, where $n$ is the power-law index of the SN outer density profile, $\rho \propto r^{-n}$. We use the ``standard" model defined in \cite{chev06}, which has $p=3.0$ and $m=0.88$ (corresponding to $n=10.3$). The time of peak flux $t_p$ is, for our purposes, equivalent to the time at which the emission at frequency $\nu_p$ has an optical depth of unity.

We can use equation \ref{eq1} to solve for $F_{\nu_p}$ and $t_p$ for a time series of radio observations at a single frequency or for $F_{\nu_p}$ and $\nu_p$ for a set of radio observations taken at multiple frequencies at a specific time. The exquisite dataset compiled by \cite{weil11} allows us to do both. \cite{chev98} and \cite{chev06} then detail how to derive physical properties of the SN and progenitor from these observables. Our derived parameters are discussed in the next section.

\begin{figure}[h]
\epsscale{1.0}
\plotone{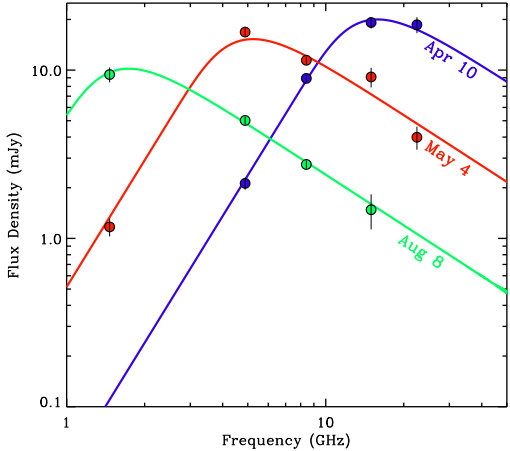}
\caption{\small Spectra for SN 1994I at three different epochs in 1994 (dates indicated on graph).  The data for each epoch are fit with a synchrotron self-absorption model with $p=3$ \citep{chev06}. The flux values are taken from \cite{weil11}.}  
\label{spec}
\end{figure}

We first considered a full spectrum of radio observations of the SN at three different epochs (Figure \ref{spec}), using the reported flux values of \cite{weil11} (green, red, and blue points). The spectra span the frequency range 1.5-22.5 GHz and are all well-fit by our synchrotron self-absorption model (solid lines). The best-fit model for each epoch was obtained by taking $t=t_p$ in Equation \ref{eq1} and minimizing a $\chi^2$ value for a range of inputs for  $\nu_p$ and $F_{\nu_p}$. The physical parameters derived from these fits are given in Table \ref{tab3} and are discussed further below. The results obtained for our second epoch, 1994 May 4, are consistent with those previously reported by \cite{stoc05} (quoted in \citealt{chev06}). 

\begin{figure*}
\epsscale{0.55}
\plotone{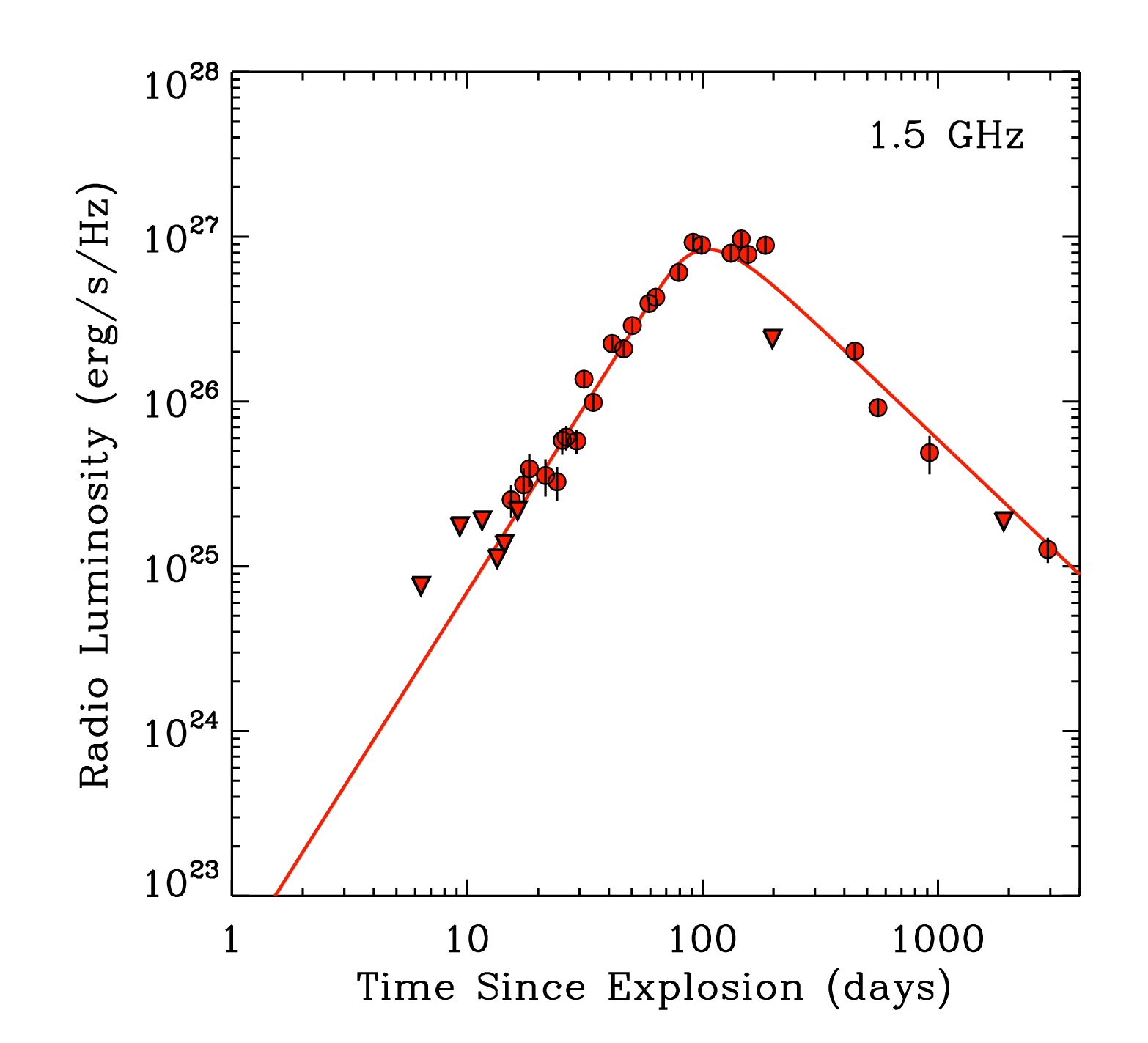}
\plotone{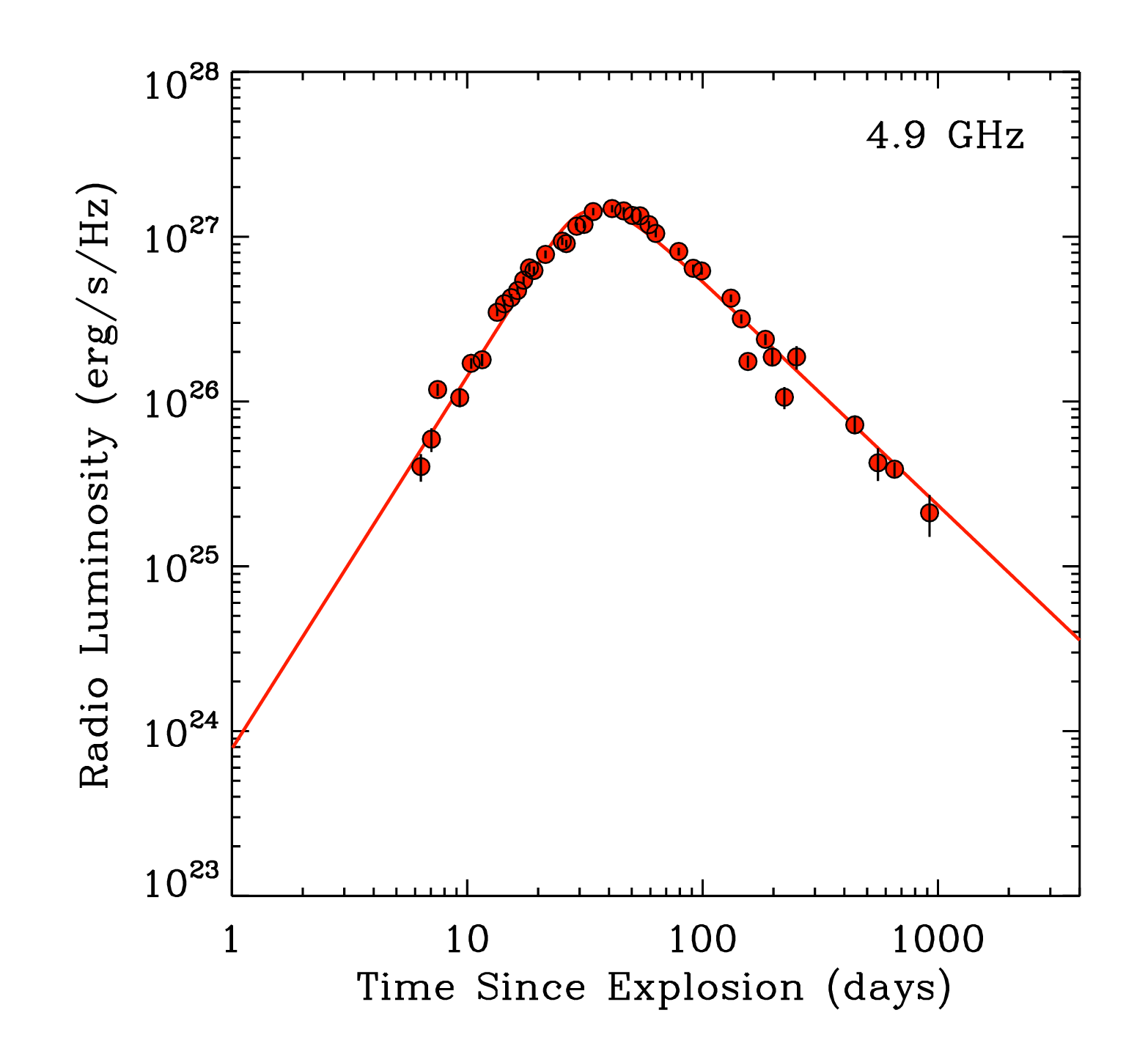}
\plotone{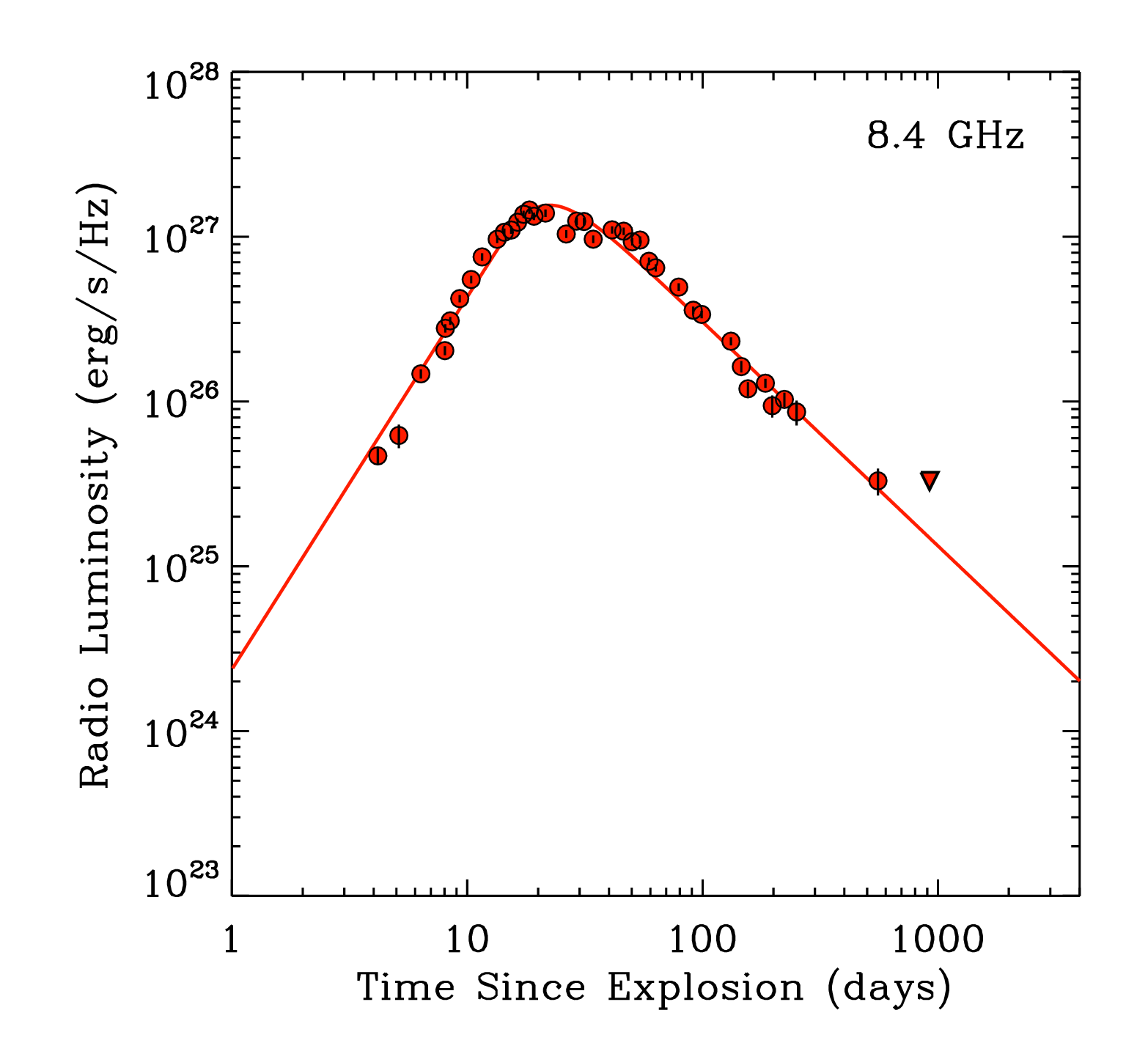}
\plotone{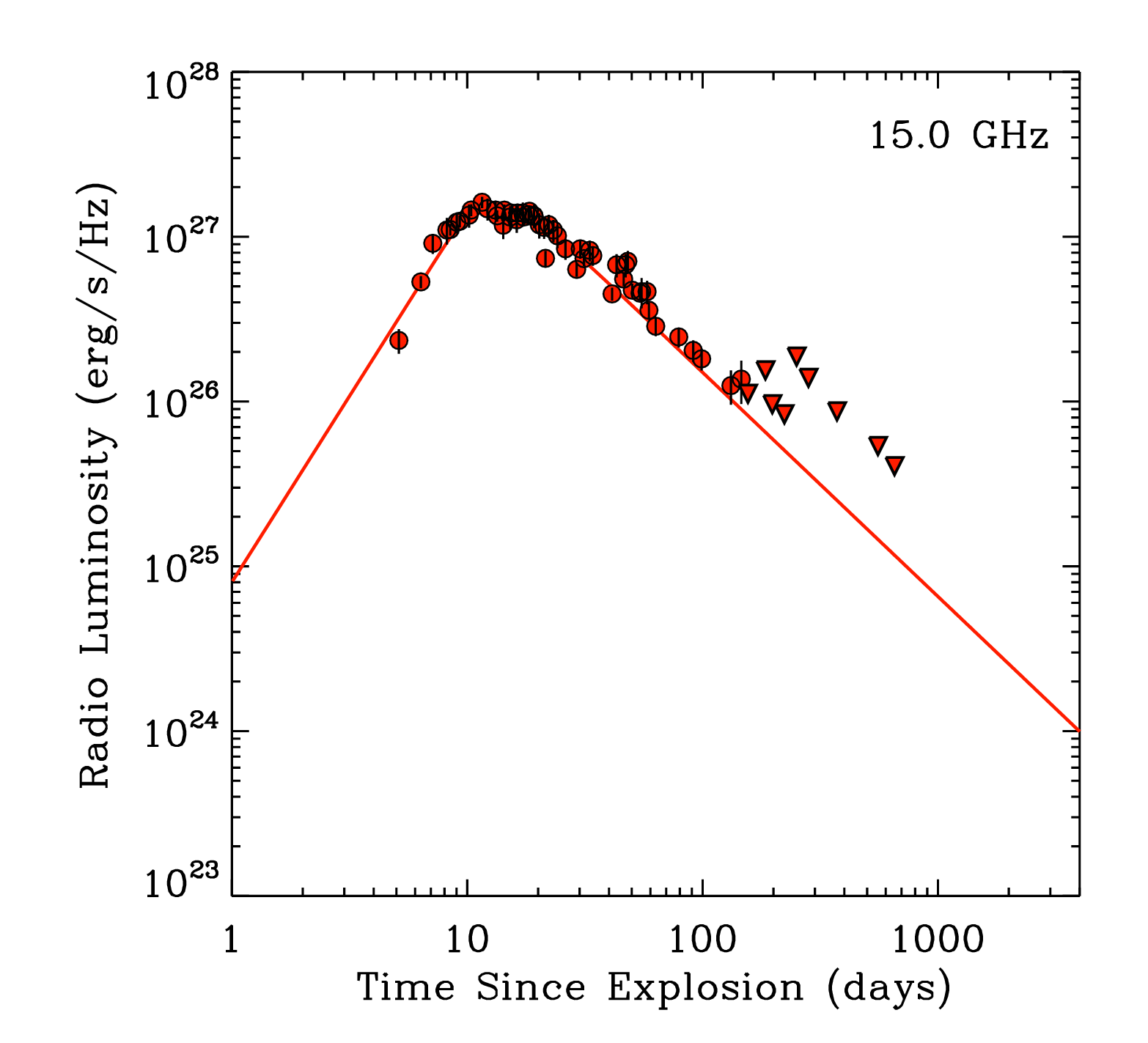}
\plotone{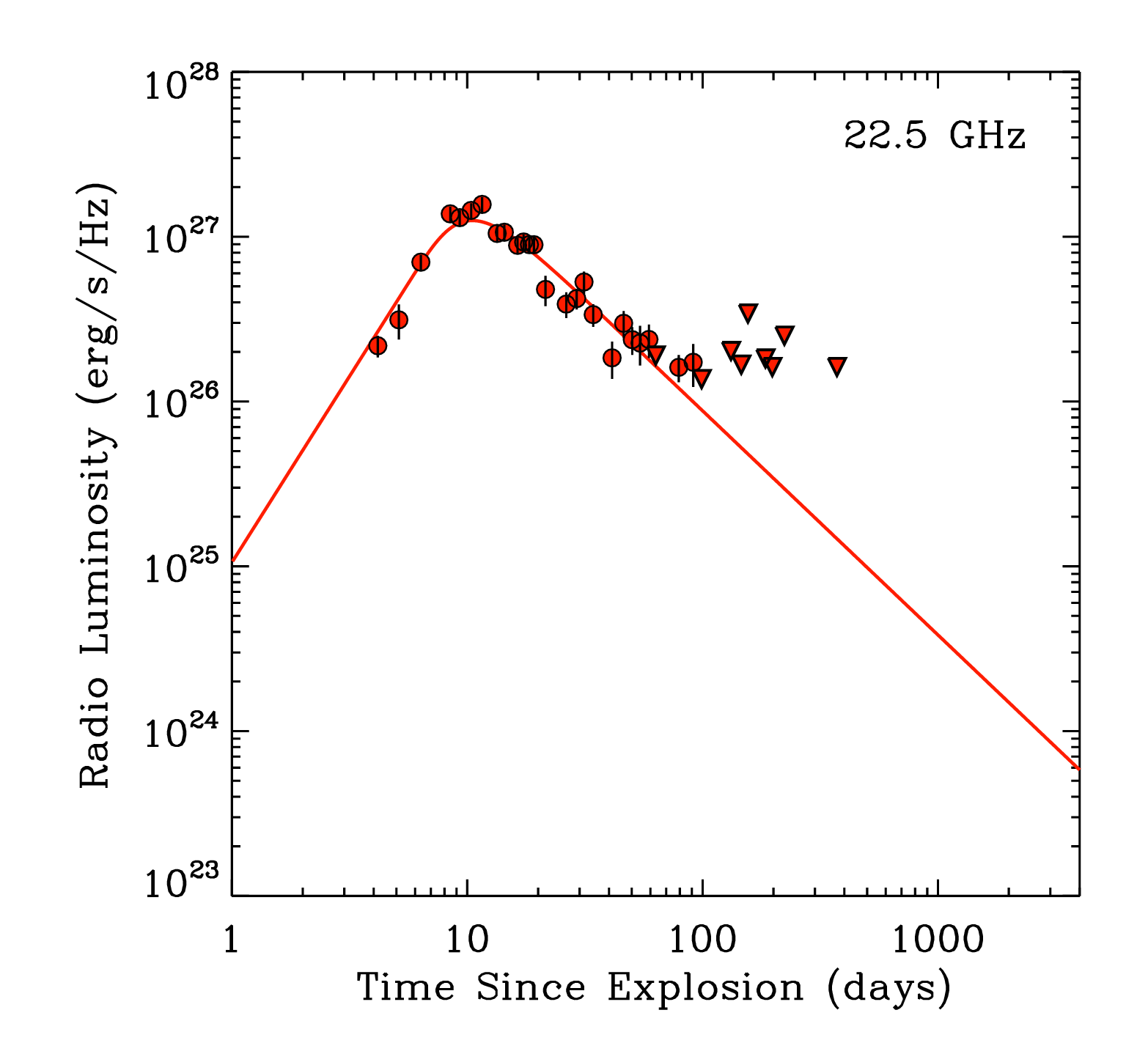}
\caption{\small Light curves of SN 1994I at five radio frequencies. The red lines are the best-fit \cite{chev06} model for each frequency. The models assume an explosion date of March 30.0 UT. The red points are flux measurements taken from \cite{weil11} and converted to luminosity assuming a distance of 8.4 Mpc for the SN. Triangles denote 3$\sigma$ upper limits.}
\label{mods}
\end{figure*}

We next considered the light curves of SN 1994I, each of which provides information about the SN system at its time of peak flux. Light curves of SN 1994I at five different frequencies are shown in Figure \ref{mods}, along with the model given in Equation \ref{eq1}. The flux values are taken from \cite{weil11}. We determine the best-fit time of peak and peak flux of each light curve by minimizing a $\chi^2$ value for a range of inputs for  $t_p$ and $F_{\nu_p}$ and taking $\nu=\nu_p$ in equation \ref{eq1}. For the 4.9 GHz light curve, we find $t_p=34.3$ days and $F_{\nu_p}=17.25$ mJy. Initially, we followed \cite{weil11} in assuming an explosion date of March 31.0 UT. However, if we allow the explosion date to vary as well and reminimize the $\chi^2$ value for each frequency, we find that the best-fit explosion date ranges from March 28.9 UT (for the 4.9 GHz observations) to March 31.1 UT (for the 22.5 GHz observations). For our modeling, we therefore assume an explosion date of March 30.0 UT, the average of these five values. This gives an overall $\chi^2=4.68$ per degree of freedom. The March 31.0 explosion date is a worse fit to the data, with an overall $\chi^2$ per degree of freedom of 5.26. The March 30.0 explosion date has also been used by \cite{mil99} in their direct spectral analysis of SN 1994I. Both March 30.0 and March 31.0 are consistent with all known early observations of the supernova and our derived physical parameters are relatively insensitive to this change in the explosion date. 

\subsection{Model Parameters}

We derive a series of physical parameters from the observables $F_{\nu_p}$ and $\nu_p$ or $t_p$ using the method outlined in \cite{chev06}. Our computed values are given in Table \ref{tab3}. In addition to the pre-explosion mass loss rate $\dot{M}$, which was previously computed by \cite{weil11} and \cite{imm02} using different methods, we calculate the blastwave radius $R$, the time-averaged blastwave velocity $\bar{v}$, the post-shock magnetic field strength $B$, and the internal energy of the shock material $E$ as functions of time. These new parameters were not previously reported in \cite{weil11}. 

Following the derivation in \cite{chev06}, the blastwave radius $R$ at time $t_p$ for a supernova described by the model in Equation 1 (with $p=3$) is given by
\begin{equation}
\begin{split}
R = & 4.0\times10^{14}\alpha^{-1/19}\left(\frac{f}{0.5}\right)^{-1/19}\left(\frac{F_{\nu_p}}{\text{mJy}}\right)^{9/19} \\
&\times \left(\frac{D}{\text{Mpc}}\right)^{18/19}\left(\frac{\nu_p}{5 \text{GHz}}\right)^{-1}\text{cm},
\end{split}
\end{equation}
and the magnetic field strength $B$ at time $t_p$ is given by
\begin{equation}
\begin{split}
B = & 1.1\alpha^{-4/19}\left(\frac{f}{0.5}\right)^{-4/19}\left(\frac{F_{\nu_p}}{\text{mJy}}\right)^{-2/19} \\
&\times \left(\frac{D}{\text{Mpc}}\right)^{-4/19}\left(\frac{\nu_p}{5 \text{ GHz}}\right)\text{G},
\end{split}
\end{equation}
where $t_p$, $\nu_p$, and $F_{\nu_p}$ are as defined in Equation \ref{eq1}, $\alpha=\epsilon_e/\epsilon_B$ is the ratio of the relativistic electron energy density to the magnetic energy density, $f$ is the filling factor (i.e. the fraction of the volume inside radius $R$ that emits in the radio), and $D$ is the distance to the supernova. Throughout this paper, we assume $\alpha=1$, $f=0.5$, and $D=8.4$ Mpc \citep{chev06,feld97}. 

Using $R$ and $B$, we can also compute the total internal energy of the radio-emitting material $E$ at each time $t_p$, which is given by
\begin{equation}
E = \frac{1}{\epsilon_B}\left(\frac{B^2}{8\pi}\right)\left(\frac{4\pi fR^3}{3}\right).
\end{equation}
For our model, we follow \cite{chev06} and assume $\epsilon_B=0.1$.

Referring back to Table \ref{tab3}, we see that $R$ and $E$ increase with time, while $B$ decreases. Our fit to the 4.9 GHz light curve ($\nu_p=4.9$) allows us to derive that $R=1.18\times10^{16}$ cm, $B=0.510$ Gauss, and $E=3.58\times10^{46}$ erg when the age of the supernova was $t=t_p=34.3$ days. We derive similar parameters from a spectrum of observations taken on May 4, 1994, when the age of the supernova was $t=35.2$ days.  This spectrum peaked at a frequency $\nu_p=4.61$ GHz. This confirms that our two methods of parameter derivation are consistent. Including both sets of fits allows us to probe the time evolution of the SN system in more detail.

We can estimate the blastwave expansion velocity using $\bar{v}=R/t$, where $R$ is the blastwave radius at time $t$ after the explosion. The blastwave velocity is a key parameter for understanding SNe. Observations of young radio supernovae (e.g., 1993J; \citealt{bar02,mar11a,mar11b})
and supernova remnants (e.g. Cassiopeia A; \citealt{and95,thor01,fes06}) indicate that the shockwaves decelerate on a timescale of decades with a roughly power-law evolution. Type Ib/c SNe have been found with a range of blastwave velocities and understanding this variation may also help illuminate the link between SNe Ic and gamma ray bursts whose jetted ejecta components decelerate on a timescale of minutes to days \citep{sod10}. We find an average expansion velocity of $\bar{v} = 35,000$ km s$^{-1}$ for SN 1994I, which decreases slightly with time over a period of months. The velocity we derive from the 4.9 GHz light curve at $t=34.3$ days is $\bar{v}=0.133c= 40,000$ km s$^{-1}$.

We next turn to the pre-explosion mass loss rate $\dot{M}$. This quantity is related to the pre-shock circumstellar density $\rho_0$ and the pre-explosion wind speed of the progenitor star, $v_w$. We can relate this to the quantities derived above if we assume that a constant fraction $\epsilon_B$ of the pre-shock energy density $\rho_0 v_w^2$ is converted to magnetic energy density at the shock boundary. In combination with the above expression for $B$, we obtain
\begin{equation}
\begin{split}
\dot{M} = &1.0\alpha^{-8/19}\left(\frac{\epsilon_B}{0.1}\right) \left(\frac{f}{0.5}\right)^{-8/19}\left(\frac{F_{\nu_p}}{\text{mJy}}\right)^{-4/19} \\
&\times \left(\frac{D}{\text{Mpc}}\right)^{-4/19} \left(\frac{\nu_p}{5 \text{ GHz}}\right) \left(\frac{t_p}{\text{10 days}}\right) \\
&\times  \left(\frac{v_w}{1000 \text{ km s}^{-1}}\right) 10^{-5} M_{\sun} \text{ yr}^{-1}.
\end{split}
\end{equation}

Note that the mass loss rate depends on the assumed pre-explosion wind speed of the progenitor, $v_w$. If we assume $v_w=1000$ km s$^{-1}$, a typical wind speed for a Wolf-Rayet star, we find an average mass loss rate of $\dot{M}=3.0 \times 10^{-5} M_{\sun}$ yr$^{-1}$. The mass loss rate derived from the 4.9 GHz light curve is $\dot{M}=2.5  \times 10^{-5} M_{\sun}$ yr$^{-1}$. These values are consistent with expectations for Wolf-Rayet mass loss \citep{chev06}. Assuming a smaller wind speed of $v_w=10$ km s$^{-1}$ decreases the average value to $\dot{M}=3.0 \times 10^{-7} M_{\sun}$ yr$^{-1}$. Our estimate is consistent with the mass loss rate $\dot{M}=1.8\times10^{-7}(v_s/$20,000 km s$^{-1})^{1.5}(v_w/10$ km s$^{-1})(T_e/$20,000 km s$^{-1})M_{\sun}$ yr$^{-1}$ reported by \cite{weil11} and is much smaller than the mass loss rate $\dot{M}=1\times10^{-5}(v_w/10 \text{ km s}^{-1})M_{\sun}$ yr$^{-1}$ obtained by \cite{imm02} from X-ray observations of SN 1994I. \cite{weil11} suggest that the discrepancy may be due to clumping of the CSM, which is not directly accounted for in their model, \cite{imm02}'s model, or ours, and they conclude that the correct mass loss rate is in between the radio and X-ray estimates. Another possibility, explored in \cite{chev06}, is that the X-ray luminosity is due to a combination of inverse Compton emission and synchrotron emission, instead of thermal emission as \cite{imm02} assume. These two emission mechanisms act to increase the X-ray luminosity for a given CSM density, meaning that the $\dot{M}$ rate give by \cite{imm02} would be biased high. As \cite{chev06} note, the X-ray observations of SN 1994I are not complete enough to fit a specific X-ray emission model, but detailed observations of the time-evolution of the X-ray luminosity of future similar SNe could be used to determine which mechanisms are responsible for the emission.

We can also use the blastwave radius $R$ and the distance to the supernova to derive the brightness temperature evolution of the SN\,1994I radio emission. Following \cite{weil11}, we use these quantities to estimate the solid angle $\Omega$ of the emitting region. ($\Omega$ could not be measured directly because SN\,1994I was not radio bright enough.) The brightness temperature is then proportional to the flux received at a given frequency divided by this solid angle, $T_B \propto F_{\nu}/\Omega$. The brightness temperature of SN\,1994I at 4.9 GHz is given as a function of time in Table \ref{tab3}. Qualitatively, we reproduce the time evolution seen by \cite{weil11}: the brightness temperature at 4.9 GHz remains roughly constant until the emission becomes optically thin at $t\approx35$ days, then drops off. This is the expected behavior for a system where synchrotron self-absorption dominates at early times. Our maximum value for $T_B$ is approximately eight times smaller than the maximum value computed by \cite{weil11}. This is largely a combination of two effects: the fact that the expansion velocity derived from our best-fit model is much faster than the constant 20,000 km s$^{-1}$ assumed by \cite{weil11}, which increases $R$, and the fact that we assume a slightly smaller distance to M51 (8.4 Mpc instead of 8.9 Mpc). Our maximum $T_B$ value is significantly below the theoretical maximum of $T_B\sim3\times10^{11}$ K \citep{kel69,rea94}, softening the upper limit on the level of external thermal absorption derived by \cite{weil11}.

\subsection{Comparisons with Previous Modeling}

Both our model and the model developed in \cite{weil02} parameterize the radio emission from supernovae according to a standard synchrotron spectrum produced via the interaction of the SN shockwave with circumstellar material (CSM). Both models make simplifying assumptions, but these assumptions differ in several important ways. Notably, \cite{weil11} assume a constant expansion velocity $v$ for the SN blastwave and a constant pre-explosion mass loss rate $\dot{M}$, while we do not. Instead, following \cite{chev98}, our model allows for deceleration of the blastwave and a non-standard (yet smooth) density profile for the explosion environment. 

In \cite{weil02}, the authors generalize their model to account for additional absorption due to distant ionized gas between the SN and the observer, the effects of free-free absorption at early times, and clumping of the CSM. This leads to the introduction of several additional free parameters, which are determined from fits to the data. In the case of SN 1994I, \cite{weil11} assume a homogeneous CSM with no clumps, just as our model does. We choose to assume only synchrotron self-absorption because this is the dominant effect at early times and this simplification allows us to explore the temporal evolution of the physical parameters of the system. We obtain a comparably good fit to the data: our model has an overall best-fit $\chi^2=4.68$ per degree of freedom (computed from the fit to light curves at all five radio frequencies observed), while \cite{weil11} report an overall $\chi^2$ value of 4.42 per degree of freedom for their model using the same light curves. 

One notable difference between our results and \cite{weil11}'s is that our derived expansion velocity is significantly higher than their assumed expansion velocity of $v=20,000$ km s$^{-1}$. We find an average expansion velocity of $\bar{v} = 35,000$ km s$^{-1}$, which appears to decrease slightly with time. This is consistent with the blastwave velocities derived for other nearby Type Ib/c SNe \citep{sod10}. Such rapid expansion velocities favor a compact progenitor over an extended stellar progenitor because compact stars have faster winds (resulting in lower-density CSM) and larger shock acceleration at the time of breakout \citep{chev10}. Our observed expansion velocity is consistent with expectations for a Wolf-Rayet progenitor, as Wolf-Rayet stars have very fast, low-density winds and strong shock acceleration in their outer layers. Combining this evidence with our computed mass loss rate, we conclude that a Wolf-Rayet star is the most likely progenitor of SN\,1994I. \cite{weil11} reach the same conclusion based on their derived mass loss rate.

\section{Radio Transients Search}
\label{sec:trans}

\subsection{Search Overview}

\begin{table*}
\center{
\caption{Bright Radio Sources Detected in M51}
\label{tab2}
{\small
\begin{tabular}{ccccc}
\hline
\hline\noalign{\smallskip}
Source    & R.A. (J2000.0)$^a$ & Dec. (J2000.0)$^a$ & Integrated 4.9 GHz  & Notes\\
          &                &                &     Flux Density (mJy)$^a$ & \\
\hline\noalign{\smallskip}
1  & 13$^{\text{h}}$29$^{\text{m}}$54$^{\text{s}}$.12 & 47$^{\circ}11'30''.33$ & variable & SN 1994I \\
2$^b$  & 13$^{\text{h}}$29$^{\text{m}}$52$^{\text{s}}$.71 & 47$^{\circ}11'42''.73$ & 1.135 $\pm$ 0.011 & Nucleus of M51$^b$ \\
3  & 13$^{\text{h}}$29$^{\text{m}}$ 51$^{\text{s}}$.57 & 47$^{\circ}12'08''.01$ & 0.946 $\pm$ 0.011 & Maddox et al. source 37 \\
4  & 13$^{\text{h}}$30$^{\text{m}}$05$^{\text{s}}$.13 & 47$^{\circ}10'35''.78$ & 4.287 $\pm$ 0.014 & Maddox et al. source 104 \\
\hline\noalign{\smallskip}
\end{tabular}

\parbox{5.5in}{
$^a$ Taken from \cite{mad07} (except for SN 1994I flux)\\
$^b$ The main point-like component of the nuclear emission.  The position and flux given are from \cite{mad07} source 53, the component detected in the largest number of epochs.  Some epochs, particularly in more compact configurations of the VLA, contain additional objects from the complicated diffuse emission near the galactic nucleus. \\
}
}
}
\end{table*}  

\begin{figure}
\epsscale{1.1}
\plotone{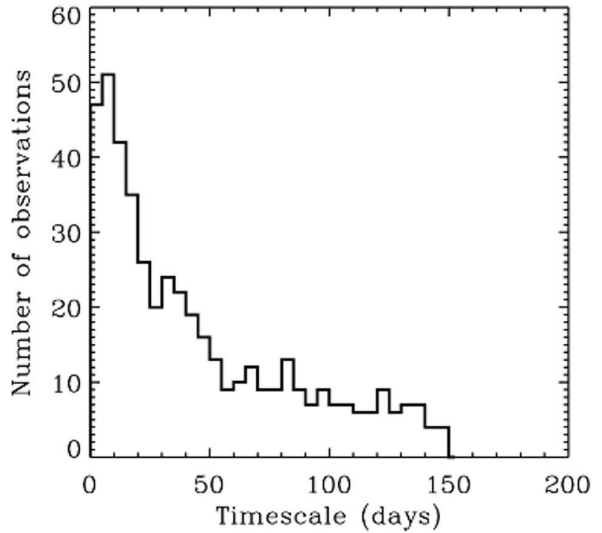}
\caption{\small Histogram showing the time separations between all possible pairs of epochs used in our transient search.  The number of observation pairs separated by a given amount of time is a rough estimate of our level of sensitivity to transients at that particular timescale.  Our sensitivity peaks at a timescale of $\sim$10 days and rapidly declines at longer timescales.}
\label{hist}
\end{figure}

\begin{figure*}
\epsscale{1.2}
\plotone{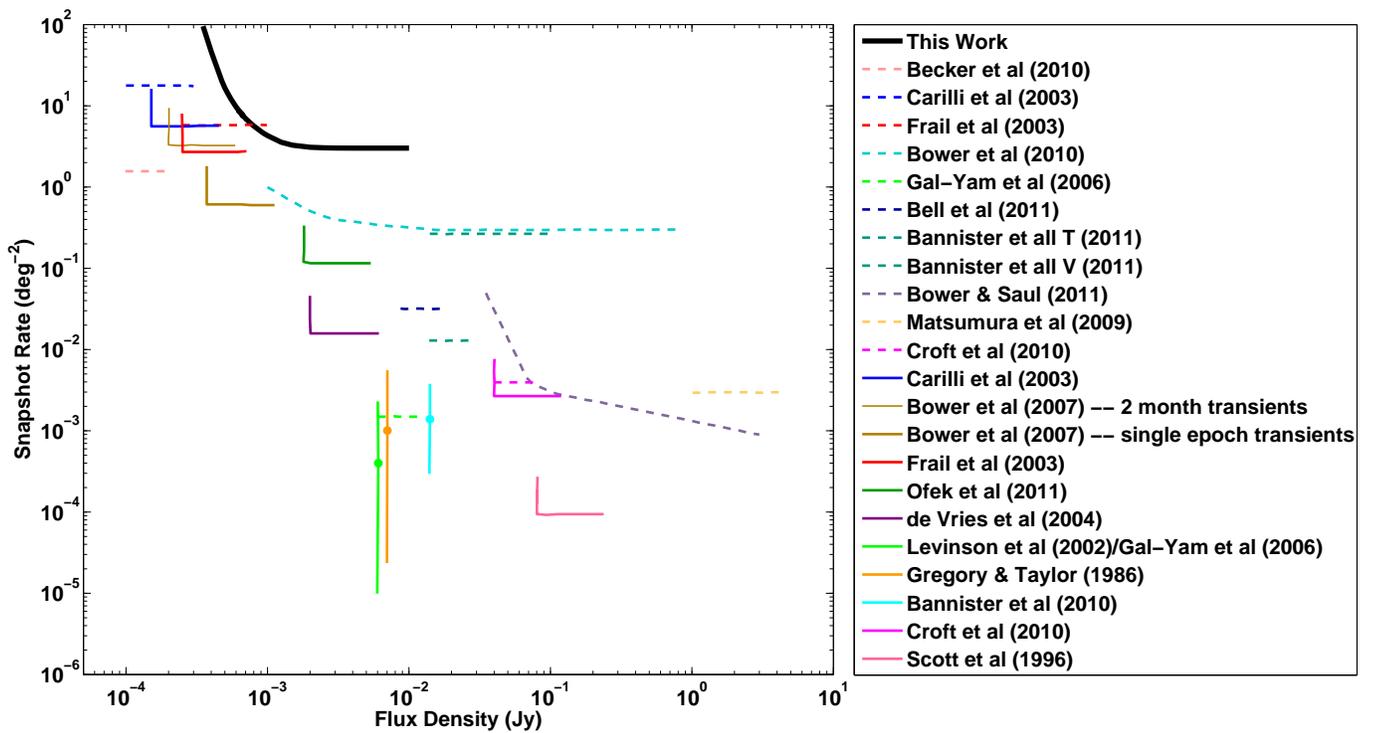}
\caption{\small Summary of current radio transient snapshot rates and upper limits taken from \cite{bel11} (dashed lines) and \cite{frail12} (solid lines).  Our 2$\sigma$ upper limit, calculating using the procedure of \cite{bel11}, is shown by the thick black line.}
\label{fig4}
\end{figure*}

We utilize this same dataset to test the feasibility of using high cadence radio supernova observations to perform additional transient science. While our pilot study only uses data from a single supernova, it nevertheless provides a unique upper limit because our search area is limited to a face-on nearby galaxy, M51. Since the distance to the galaxy is known, we are able to set an upper limit on the rate of transients in M51 above a specific luminosity threshold. After confirming the quality of our maps by comparing our observed SN 1994I 4.9 GHz flux values to those reported by \cite{weil11}, we searched these maps for other transient objects. We used 31 epochs for our radio transient search, each 9-20 minutes in duration with typical sensitivities of 60-100 $\mu$Jy beam$^{-1}$ rms. The detection threshold was set at 5$\sigma$ or 6$\sigma$, depending on image quality, resulting in an average central detection threshold of 430 $\mu$Jy. The observations are logarithmically spaced, with a time difference of 1-10 days between consecutive epochs.  The non-constant spacing between epochs makes us varyingly sensitive to transients on different characteristic timescales, ranging from 10 minutes to 5 months.  However, since we have many pairs of observations separated by timescales of a few days, and only one pair separated by 5 months, we note that our sensitivity to longer-duration transients is low (see Figure \ref{hist}). In particular, we might be expected to miss most transients occurring at timescales longer than about 10 days (the maximum separation between two consecutive epochs). 

We generated a list of radio sources in each epoch using SAD, as described in Section \ref{sec:obs}.  This list was compared with the catalog of radio sources in M51 compiled by \cite{mad07}, which was created using much deeper radio observations of M51 at 1.425 GHz and 4.860 GHz.  The total on-source integration time for their 4.86 GHz data was 22.5 hours and their images reached a sensitivity of 11.7 $\mu$Jy beam$^{-1}$ rms.  SN 1994I was detected in all of our epochs, as was emission from the nucleus of M51.  Two other bright sources were also regularly detected, and therefore classified as non-transients. Both of these sources appear in the \cite{mad07} catalog (sources 37 and 104). The positions and fluxes of these four objects are given in Table \ref{tab2}.  In addition to these four steady sources, we detect \cite{mad07} source 107 in five epochs. Our clearest detection is on 1994 August 8, when we find its integrated flux to be 0.44 $\pm$ 0.10 mJy. \cite{mad07} report a flux of $0.482 \pm 0.020$ mJy. The fact that we don't detect this object in all of our epochs may indicate that its emission is variable, but it may also be because this flux is near the limit of our sensitivity. \cite{mad07} classify this source as an X-ray binary.

In addition to these five sources that appear in multiple epochs of data, we initially detected 99 candidate transient objects that each only appear in an individual epoch. This constrains the characteristic timescales of the candidates to between 9 minutes (the minimum epoch duration) and 10 days (the maximum separation between consecutive epochs).  A significant fraction of these candidates were discarded as imaging defects based on a visual inspection of the data, leaving us with a list of 30 possible transients.  Two of our transient candidates were also detected by \cite{mad07}, who found them to have flux densities of $264 \pm 14$ $\mu$Jy beam$^{-1}$ and $443 \pm 12$ $\mu$Jy beam$^{-1}$, respectively (\citealt{mad07} Table 2, sources 3 and 97). We obtained flux densities of $341 \pm 61$ $\mu$Jy beam$^{-1}$ for source 3 and $378 \pm 57$ $\mu$Jy beam$^{-1}$ for source 97. Both of these objects were only detected on 1994 August 8, which was the deepest observation collected when the VLA was in B configuration. The fact that we only detected these two objects in a single epoch of data may be because their fluxes are below our average sensitivity limit, but we cannot rule out variability.

We further refined our list of transient candidates by manual reimaging of epochs containing one or more candidates.  The observations for each epoch contain two IFs and two polarizations, so we reimaged the epochs containing one or more candidates, taking each IF and each Stokes parameter separately.  Objects not detected in all four of the resulting images were discarded as false detections, eliminating all but four candidates.  All of the epochs in our transient search also contain data taken at other radio frequencies, so we reduced this data to see if these remaining candidate transients were detected in these frequencies as well.  We found that none of our four candidates was detected at 8.4 GHz or 1.5 GHz. Our typical detection thresholds at these frequencies were 400 $\mu$Jy beam$^{-1}$ and 600 $\mu$Jy beam$^{-1}$ respectively. As three of the candidates were marginal detections in a single epoch and the fourth was detected near the location of the nucleus of M51 under poor observing conditions, we conclude that none of our candidates is a real transient.

\subsection{Radio Transient Population in M51}
Our non-detection of transient objects in M51 allows us to place an upper limit on the transient population of this galaxy.  Our mean central detection threshold of 430 $\mu$Jy corresponds to a luminosity of 3.7 x 10$^{25}$ erg s$^{-1}$ Hz$^{-1}$ at the distance of M51 (8.4 Mpc).  This is comparable to the luminosity of Cassiopeia A, the most luminous SN remnant in the Milky Way \citep{rey02}. We can therefore set a limit on the occurrence rate of all types of transients more luminous than this, including radio supernovae and X-ray binaries. The sensitivity of our images decreases with increasing radius, so to determine the observed rate of transients above a given flux density, we must first determine the area in each image within which we can detect objects of at least the chosen brightness (see \citealt{bow07}).  For the VLA, the half power radius at 4.9 GHz is equal to 4.5' (as determined above), and we compute the viable area in each image assuming a Gaussian decline in sensitivity. We then add up the viable areas from all of the images to obtain the total effective area $N$. The transient density (also called the snapshot rate) is the number of transients observed divided by $N$.  Since we did not detect any transients, we follow the procedure of \cite{bel11} to determine a 2$\sigma$ upper limit for the snapshot rate.  For zero detections, they define the snapshot rate $\rho$ using the equation
\begin{equation}
P(n) = e^{-\rho N},
\label{eq:rate}
\end{equation}
where $P(n)=0.05$ corresponds to the 2$\sigma$ upper limit with a confidence level of 95$\%$.  Thus, we can determine the upper limit for the snapshot rate of transients at any flux density by simply finding $N$ for a given flux density cutoff and solving for $\rho$ in Equation \ref{eq:rate}.  For example, using this procedure, we find an upper limit of 17 deg$^{-2}$ for the snapshot rate for transients of at least 0.5 mJy (equivalent to a luminosity of $4.2 \times 10^{25}$ erg s$^{-1}$ Hz$^{-1}$ at the distance of M51).  The current best estimates of the radio transient snapshot rate for various flux cutoffs are given in Figure \ref{fig4}, with our upper limit included as the solid black line (figure adapted from \citealt{bel11}). While this limit is less restrictive than other quoted works, it is also the only limit for a spatial region covering a nearby face-on spiral galaxy. We therefore expect to observe more nearby transients, so the average luminosity of transients at a given flux cutoff will be lower. Since brighter transients are rarer, this causes the snapshot rate of our field to be higher than the snapshot rate in the empty spatial regions covered by the other surveys and makes our upper limit significant. It is also possible to calculate an actual transient rate (the number of transients observed per deg$^2$ per year), but given the uncertainty in the characteristic timescales of the transients to which we would be sensitive (10 minutes to 10 days, as mentioned above), this is not a useful constraint. Again, Figure \ref{fig4} understates the value of our result because unlike the other quoted upper limits, we can feasibly convert our flux limit to a luminosity limit, assuming that most of the sources in our field are associated with M51. This allows us to place limits on the occurrence rate of specific classes of transient objects within M51. A larger study expanded to additional nearby galaxies could make broader claims about the occurrence rates of such transients in spiral galaxies more generally.

\section{Conclusions}
\label{sec:conc}
We presented a revised model for the radio emission from SN 1994I, which fits the data well. Our model allows us to calculate a number of interesting physical properties of the system, including the average expansion velocity of the blastwave as a function of time, the magnetic field strength, and the pre-explosion mass loss rate. The physical parameters of the system calculated from this model are consistent with a Wolf-Rayet progenitor for the SN. Our mass loss rate is consistent with the rate previously published by \cite{weil11} to within errors and assumptions. We also presented the results of a search for radio transients in M51 using archival 4.9 GHz VLA data, supplemented by observations at 1.5 GHz and 8.4 GHz.  We detected no transients in 31 epochs spanning a five month period, setting an upper limit of 17 deg$^{-2}$ on the transient density above 0.5 mJy. Since the distance to M51 is known, we can convert this to a luminosity limit on the density of all types of transients within the galaxy brighter than $L\sim4\times10^{25}$ erg s$^{-1}$ Hz$^{-1}$. This is consistent with previously published upper limits and transient detection rates, and still allows for the transient rate of M51 to be higher than that observed in empty portions of the sky.  Our pilot study shows that radio transient programs focused around a smaller field of view at high resolution are feasible with current facilities, particularly in regions with numerous existing archival observations. In particular, our work suggests a potentially exciting new use for datasets originally collected to study the evolution of radio supernovae. The recently upgraded VLA is a perfect instrument to conduct a larger high resolution radio transients search, as has already been suggested by \cite{frail12}. The proposed VLA Sky Survey is expected to produce significant advances in radio transient science. The search for transients in nearby galaxies holds the potential to be especially scientifically interesting due to the higher probability of the presence of objects that could generate detectable transient radio emission, such as radio supernovae or X-ray binaries.

\acknowledgments{}
\section{Acknowledgments} 

Thanks to Jonathan McDowell and Marie Machacek for helpful comments on early drafts of this paper.

This work is supported in part by the National Science Foundation Research Experiences for Undergraduates (REU) and Department of Defense Awards to Stimulate and Support Undergraduate Research Experiences (ASSURE) programs under Grant no. 0754568 and by the Smithsonian Institution.

The National Radio Astronomy Observatory is a facility of the National Science Foundation operated under cooperative agreement by Associated Universities, Inc. 

This research has made use of SAOImage DS9, developed by Smithsonian Astrophysical Observatory.

\bibliography{paper}

\end{document}